# Electron Spin-Relaxation Times of Phosphorus Donors in Silicon

*A.M. Tyryshkin[1], S.A. Lyon[1], A.V. Astashkin[2], A.M. Raitsimring[2]*

[1]Department of Electrical Engineering, Princeton University, Princeton, NJ 08544

[2]Department of Chemistry, University of Arizona, Tucson, AZ 85721

## Abstract

Pulsed electron paramagnetic resonance measurements of donor electron spins in natural phosphorus-doped silicon (Si:P) and isotopically-purified $^{28}$Si:P show a strongly temperature-dependent longitudinal relaxation time, $T_1$, due to an Orbach process with $\Delta E = 126$ K. The 2-pulse echo decay is exponential in $^{28}$Si:P, with the transverse relaxation (decoherence) time, $T_2$, controlled by the Orbach process above ~12 K and by instantaneous diffusion at lower temperatures. Spin echo experiments with varying pulse turning angles show that the intrinsic $T_2$ of an isolated spin in $^{28}$Si:P is ~60 ms at 7 K.

Spin-based quantum information processing has recently occupied the attention of a considerable number of researchers. In the past few years several physical implementations of a spin quantum computer have been proposed, including nuclear magnetic resonance (NMR) of small molecules [1], fullerene structures with trapped nitrogen and phosphorus atoms [2], shallow donors and quantum dots in semiconductors [3], and electrons in multilayer heterostructures [4]. Electron or nuclear spins, and combinations of the two have been proposed as qubits (quantum bits). A seven-qubit NMR quantum computer has been demonstrated experimentally [5].

A long coherence time is the feature which makes spins attractive as qubits. Decoherence limits the number of elementary gate operations which can be performed, and while error-correction algorithms partially circumvent the effects of decoherence, it has been estimated that still at least $10^4$ operations must be performed within a coherence time [6]. Electrons bound to shallow donors in silicon were reported over 40 years ago to have long spin relaxation times [7,8], with phosphorus donors (Si:P) being the most extensively studied. At low donor concentrations ($< 10^{16}$ P/cm$^3$), the longitudinal (spin-lattice) relaxation time, $T_1$, in Si:P is found to be independent of the phosphorus concentration, and it varies from microseconds at 20 K to thousands of seconds at 2 K [7,9]. The more relevant time for quantum information processing is the transverse relaxation time, $T_2$, which if measured for isolated spins can be associated with the decoherence time. In isotopically-purified $^{28}$Si:P, $T_2 \sim 500$ μs was previously reported [8]. It has been argued that maintaining the fidelity of gate operations requires that they be no shorter than a few microseconds [10], thus implying $<10^3$ operations in the 500 μs.

Here we report a detailed new study of $T_1$ and $T_2$ for P donors in $^{28}$Si and natural Si over the temperature range 7-20 K using pulsed electron paramagnetic resonance (EPR). The key new result is that the intrinsic $T_2$ (of an isolated spin) in $^{28}$Si:P is determined to be at least 2 orders of magnitude longer than the earlier reports. After taking into account the dipole-dipole interactions of neighboring spins (instantaneous diffusion), and overcoming some instrumental limitations, we have measured the intrinsic $T_2$ to be approximately 60 ms for a phosphorus donor in $^{28}$Si at 7 K. These results show that electron spins in doped, isotopically-pure $^{28}$Si have long enough coherence times to be considered for quantum computation, including reasonable estimates of the time required for elementary gate operations.

Four silicon samples with different phosphorus concentrations were studied: two samples with 0.8·10$^{15}$ and 1.7·10$^{16}$ P/cm$^3$ in natural silicon (referred in the text as "Si:P-10$^{15}$" and "Si:P-

$10^{16}$", respectively), and two samples with $0.87 \cdot 10^{15}$ and $1.6 \cdot 10^{16}$ P/cm$^3$ in isotopically-purified $^{28}$Si (referred as "$^{28}$Si:P-$10^{15}$" and "$^{28}$Si:P-$10^{16}$", respectively). Pulsed EPR measurements at X-band (9.7 GHz) were done on a Bruker Elexsys 580 EPR spectrometer using a flex-line resonator (EN-4118MD4). Measurements at C-band (4.7 GHz) and K$_u$-band (16.3 GHz) were done on a multifrequency pulsed-EPR spectrometer described elsewhere [11]. Low temperatures were achieved using helium-flow cryostats (Oxford CF935). Temperature was controlled with a precision of better than 0.05 K using calibrated temperature sensors (Lakeshore Cernox CX-1050-SD) and Oxford ITC503 temperature controllers. This precision was required because of the strong temperature dependence of the relaxation times (for example, $T_1$ varies by 5 orders of magnitude between 7 K and 20 K).

An inversion recovery experiment and a 2-pulse echo experiment were used to measure the longitudinal, $T_1$, and transverse, $T_2$, relaxation times, respectively. In the inversion recovery experiment ($\pi - T - \pi/2 - \tau - \pi - \tau -$ echo) the delay, T, after the first (inversion) pulse was varied while $\tau$ was kept constant, and the amplitude of the primary echo signal formed by the second and third pulses was measured. In the 2-pulse echo experiment ($\pi/2 - \tau - \pi - \tau -$ echo), the amplitude of the echo signal was measured as a function of the inter-pulse delay, $\tau$. The $\pi/2$ and $\pi$ pulses in both experiments were 16 and 32 ns, respectively, allowing full excitation of the EPR lines.

The EPR spectrum of phosphorus donors in silicon consists of two lines centered at g = 1.9992 and split by 41.94 G due to the hyperfine interaction with the $^{31}$P nucleus. At 15 K and below the EPR lines have a Gaussian shape, both in Si:P and $^{28}$Si:P. This indicates that an inhomogeneous line broadening due to hyperfine interaction with $^{29}$Si nuclei is the main source of the spectral linewidth in both samples. From comparing the EPR linewidths in Si:P-$10^{16}$ and $^{28}$Si:P-$10^{16}$ ($\Delta B$ = 2.5 G and 0.08 G, respectively [12]), we estimate that the residual $^{29}$Si concentration in $^{28}$Si:P does not exceed 50 ppm. Most of the relaxation measurements were done on the high field component of the donor EPR spectrum, though identical relaxation decays were seen at the low field component.

The longitudinal relaxation time in Si:P has already been studied in detail. A strong temperature dependence observed in the 2-20 K range was suggested to arise from an Orbach relaxation process. An energy splitting, $\Delta E$ = 123 K, was derived from fast-passage experiments using continuous wave (cw) EPR and shown to be a measure of the valley-orbit energy splitting of the P donor [9]. Below 2 K $T_1$ was observed to stay approximately constant, limited by direct phonon processes [7].

In the inversion-recovery experiments both the Si:P and $^{28}$Si:P samples showed mono-exponential decays, and $T_1$ was obtained by fitting to a simple exponential. The temperature dependence of $1/T_1$ obtained for $^{28}$Si:P-$10^{16}$ at 9.7 GHz is shown with filled circles in Fig. 1a. An identical temperature dependence was seen for the other samples. This confirms the previous observation that $T_1$ is independent of P concentration (at $< 10^{16}$ P/cm$^3$), and also shows that $T_1$ does not change upon $^{29}$Si-depletion [7,8]. Quite remarkably, $T_1$ varies by 5 orders of magnitude over the temperature interval 7-20 K. In Arrhenius coordinates, $\log(1/T_1)$ *vs.* $1/T$, the dependence is linear which is consistent with an Orbach mechanism dominating the relaxation process at these temperatures ($1/T_1 \propto \exp(-\Delta E/kT)$). From the slope in this plot the energy gap to the excited state involved in the relaxation process is found to be $\Delta E = 126.1 \pm 0.5$ K, in good agreement with $\Delta E = 123$ K derived from the cw measurements [9].

The measured dependence of $T_1$ extends over a significant temperature range, 7-20 K, and as a consequence a good fit of the experimental data cannot be obtained assuming a two-phonon Raman process with a power-law dependence of $1/T_1 \propto T^7$ [13]. In principle, the Raman and Orbach mechanisms may contribute jointly into the $T_1$ relaxation processes. To evaluate a possible contribution of a Raman process experiments were done at two additional microwave frequencies, $\nu_{mw} = 4.7$ and 16.3 GHz. The Raman process is expected to show a quadratic frequency dependence ($1/T_1 \sim \nu_{mw}^2$) [13], and its contribution should differ by an order of magnitude at these two frequencies. The $T_1$ data for Si:P-$10^{16}$ included in Fig. 1a do not show a significant variation between 4.7 and 16.3 GHz (the slight differences most likely arise from offsets in the temperature calibrations). Thus, we may conclude that the contribution of the Raman process to $T_1$ is negligible, and the $T_1$ relaxation in Si:P is dominated by the Orbach process over the temperature range 7-20 K.

No detailed temperature dependence study of $T_2$ has been reported, but data from several authors show that $T_2$ also varies over a wide range, from 0.3 µs at 25 K to 0.6 ms at 1.6 K [14,15]. It was shown that $T_2$ saturates at considerably higher temperatures than $T_1$ and at considerably lower values (i.e. $T_2 = 0.6$ ms *vs.* $T_1 = 3 \cdot 10^3$ s at 1.2 K) [7,15]. The mechanism of this saturation has not been explained. In natural Si:P (4.7% $^{29}$Si), non-exponential electron spin echo (ESE) decays were observed, best described by V(t) = $\exp(-t/T_2 - t^3/T_S^3)$ [8,15]. The cubic exponential term was explained as being caused by spin diffusion in the nuclear $^{29}$Si system [15]. This suggests that in isotopically-purified $^{28}$Si the term with $T_S$ should vanish and pure exponential decays, presumably very long, should be observed. However, in their original work Gordon and Bowers reported only about a factor of 2 increase in $T_2$ for $^{28}$Si:P over natural Si:P at 1.4 K [8].

The $T_2$ relaxation decays in the present study were measured using a conventional 2-pulse ESE sequence. At long inter-pulse delay $\tau$ (> 0.3 ms), we observed significant fluctuations in the phase of the detected echo signal with respect to the phase of the microwave source. These fluctuations originate from non-ideal characteristics of the spectrometer and may result from phase instability of the microwave source or from fluctuations in the external magnetic field during the 2-pulse experiment (these two types of fluctuation are indistinguishable in their effect on the relative phase of the echo signal). Because of this instrumental phase noise, repetitive summation of the echo signal using a conventional quadrature receiver (where the echo intensity is detected with respect to the phase of the microwave source) resulted in distorted echo decays, with a strongly non-exponential behavior (see trace labeled "quadrature detection" in Fig. 2). To avoid this instrumental problem, we implemented a different approach of signal accumulation consisting of repetitive summation of the *magnitude* of the echo signal calculated as [(in-phase)$^2$ + (out-of-phase)$^2$]$^{1/2}$, where "in-phase" and "out-of-phase" are the signals from the two channels of the quadrature receiver. As a result, nearly exponential decays were recovered in $^{28}$Si:P over the entire range of $\tau$ (labeled "magnitude detection" in Fig. 2) [16].

Figure 2 compares the 2-pulse ESE decays in isotopically-purified $^{28}$Si:P-10$^{15}$ and in natural Si:P-10$^{15}$ (both measured with the magnitude detection approach). While the decay is nearly exponential in $^{28}$Si:P, it is non-exponential in the natural Si:P, well-described by $\exp(-t/T_2 - t^3/T_S^3)$ [15]. With B$_0$ oriented along a (100) axis of the Si crystal, $T_S$ = 0.63 ms is estimated from the fit, and this $T_S$ is found to be temperature independent over the range 7-12 K. A shorter $T_S$ (0.36 ms) was reported in the earlier study by Chiba and Hirai [15] but in that work the crystal was oriented differently. We found that $T_S$ is orientation-dependent in natural Si:P, and the longest $T_S$ is seen for a (100) orientation. The orientation dependence of $T_S$ will be discussed in more detail elsewhere [17].

The linear term in the exponential, $T_2$, can also be derived from the fit $\exp(-t/T_2 - t^3/T_S^3)$ of the decays in natural Si:P, and it is found to be the same as $T_2$ in isotopically-purified $^{28}$Si:P, measured at the same temperature and P concentration [18]. Thus, it is the presence of the cubic $T_S$ term which makes the difference between 2-pulse echo decays in Si:P and $^{28}$Si:P. It is seen in Fig. 2 that in Si:P the $T_S$ term dominates at $\tau > 0.2$ ms and results in complete suppression of the 2-pulse echo signal within first 0.5 ms.

Figure 1b shows the temperature dependence of $T_2$ in two $^{29}$Si-depleted samples, $^{28}$Si:P-$10^{16}$ and $^{28}$Si:P-$10^{15}$ at 9.7 GHz. The measurements at 4.7 and 16.3 GHz gave similar $T_2$ dependences (not shown). For comparison, the $T_1$ data are also included in Fig. 1b. Two temperature ranges are clearly seen in the $T_2$ data. At high temperatures, 12-20 K, $T_2$ follows closely and nearly coincides with $T_1$. Apparently, the $T_1$ relaxation process (the Orbach process) makes the major contribution to $T_2$ over this temperature range. However, at temperatures lower than 12 K in $^{28}$Si:P-$10^{16}$ and lower than 10 K in $^{28}$Si:P-$10^{15}$, $T_2$ diverges from $T_1$. While $T_1$ continues to grow, approaching to $3 \cdot 10^3$ s at 1.2 K [7], $T_2$ becomes temperature-independent and saturates at 0.27 and 2.8 ms in $^{28}$Si:P-$10^{16}$ and $^{28}$Si:P-$10^{15}$, respectively. The fact that the limiting value of $T_2$ is greater in the sample with smaller P concentration suggests that $T_2$ at low temperatures is mostly determined by the dipole-dipole interaction between neighboring P-donors via instantaneous diffusion [19]. This effect is known to be temperature-independent, given that the $T_1$ of the spins involved is long compared to the total duration of the 2-pulse experiment [20].

To test this hypothesis, 2-pulse echo experiments ($\pi/2 - \tau - \theta_2 - \tau -$ echo) with a variable rotation angle, $\theta_2$, of the second microwave pulse were performed. The results obtained at 6.9 and 8.1 K are plotted in Fig. 3. It is seen that $1/T_2$ varies linearly with $\sin^2(\theta_2/2)$ indicating that, indeed, instantaneous diffusion contributes significantly to the observed $T_2$ relaxation rates at these temperatures. Linear fits were obtained assuming the same slope for both data sets, and the resulting slope was $(3.2 \pm 0.2) \cdot 10^2$ s$^{-1}$, which is close to $3.5 \cdot 10^2$ s$^{-1}$ as expected for a homogeneous P-distribution at $0.87 \cdot 10^{15}$ P/cm$^3$ [21]. Extrapolation to $\theta_2 = 0$ (to eliminate the contribution of instantaneous diffusion) gives $1/T_2 = 0.072 \pm 0.008$ and $0.016 \pm 0.007$ ms$^{-1}$ at 8.1 K and 6.9 K, respectively. The corresponding intrinsic $T_2$ values (of the isolated donor-electron spins in $^{28}$Si:P) are $14 \pm 2$ ms at 8.1 K and 62 (+50/−20) ms at 6.9 K and approach the $T_1$ values at these temperatures (18.5 and 280 ms, respectively).

In summary, we demonstrate that the observed low-temperature $T_2$ relaxation time in $^{28}$Si:P is controlled by at least two processes, the $T_1$ relaxation mechanism and instantaneous diffusion, and that the intrinsic $T_2$ of isolated P donors in Si approaches $T_1$ at temperatures as low as 6.9 K. At this temperature we find that the intrinsic $T_2 < T_1$, but it is not yet clear whether this difference is just experimental error or if it is indicative of a new decoherence mechanism. A resolution of this question and the possibility of finding even longer intrinsic $T_2$'s in $^{28}$Si:P at lower temperatures is currently limited by signal-to-noise considerations arising from the necessity to use magnitude detection at donor concentrations below $10^{14}$ P/cm$^3$.

The observation of an extremely long $T_2$ has important implications for efforts to use bound electron spins in Si as qubits for quantum information processing. Decoherence is inevitable, but it has been shown that quantum error correction can make information processing possible if the error rate is sufficiently low to allow at least $10^4$ elementary gate operations to be performed [6]. At the same time, these operations must be performed with great precision, of the order of 1 part in $10^4$. For such precision to be possible it has been estimated that microwave pulses to perform single-qubit operations must be at least a few microseconds in length [10]. Thus, the electron spin decoherence times must be at least a few tens of milliseconds. We have now directly shown that electrons bound to phosphorus donors have a long enough $T_2$ (> 60 ms) to satisfy these requirements.

To achieve this long electron spin decoherence time (in any candidate spin system), some conditions must be satisfied. One is that the electrons must be kept well separated. We have found that the magnetic dipole-dipole interaction is the dominant decoherence process for donor densities above about $10^{13}$/cm$^3$ and temperatures below 7 K. To obtain decoherence times of 60 ms or longer, the bound electrons must be at least several tenths of a micron apart. Another condition is that the electrons should be strongly bound. The electrons bound to phosphorus donors are only able to show long decoherence times because the lowest excited state is far away and the Orbach relaxation process can be frozen out. Quantum dots or other bound systems will either need to be no larger than a few nm in size, or much lower temperatures will be necessary to freeze out the Orbach process.

The work at Princeton was supported in part by the U.S. Army Research Office and the Advanced Research and Development Activity under Contract No. DAAD19-02-1-0040 and the Defense Advanced Research Projects Agency's SPINS Program through Los Alamos National Lab. The development of the multifrequency spectrometers and work at the University of Arizona was supported in part by the National Science Foundation under grants DBI-9604939 and BIR-9224431.# References

[1]     N. A. Gershenfeld and I. L. Chuang, Science **275**, 350 (1997).

[2]     W. Harneit, Phys. Rev. A **65**, 032322 (2002).

[3]     B. E. Kane, Nature **393**, 133 (1998); D. Loss and D. P. DiVincenzo, Phys. Rev. A **57**, 120 (1998).

[4]     R. Vrijen, et al., Phys. Rev. A **62**, 012306 (2000).

[5]     D. G. Cory, et al., Fortschritte Phys.-Prog. Phys. **48**, 875 (2000).

[6]     D. P. DiVincenzo, Fortschritte Phys.-Prog. Phys. **48**, 771 (2000).

[7]     G. Feher and E. A. Gere, Phys. Rev. **114**, 1245 (1959).

[8]     J. P. Gordon and K. D. Bowers, Phys. Rev. Lett. **1**, 368 (1958).

[9]     T. G. Castner, Phys. Rev. Lett. **8**, 13 (1962).

[10]   M. Friesen, et al., e-print cond-mat/0204035 (2002).

[11]   P. P. Borbat and A. M. Raitsimring, in *The 36th Rocky Mountain Conference on Analytical Chemistry*, Denver, CO, July 31-August 5, 94 (1994); A. V. Astashkin, et al., in *The 40th Rocky Mountain Conference on Analytical Chemistry*, Denver, CO, July 26–30, add. materials (1998).

[12]   Measurements were done to ensure no distortion to the EPR lineshape. Experimental conditions: microwave power = 0.2 µW, modulation amplitude = 10 mG, modulation frequency = 30 kHz.

[13]   A. Abragam and B. Bleaney, *Electron paramagnetic resonance of transition ions* (Dover Publications, NY, 1986).8

Figure Captions.

Fig. 1. Temperature dependence of the longitudinal, $T_1$, and transverse, $T_2$, relaxation times in (100)-oriented Si:P. (a) $T_1$ is shown at three microwave frequencies: in isotopically-purified $^{28}$Si:P-$10^{16}$ at 9.7 GHz (●), and in natural Si:P-$10^{16}$ at 16.3 GHz (■) and 4.7 GHz (▼). (b) $T_2$ is shown (○) for two samples, $^{28}$Si:P-$10^{16}$ and $^{28}$Si:P-$10^{15}$, at 9.7 GHz. For reference, the $T_1$ dependence for $^{28}$Si:P-$10^{16}$ (●) is reproduced from (a).

Fig. 2. Semilog plot of the 2-pulse ESE decay as a function of the interpulse delay, $\tau$, for Si:P-$10^{15}$ and $^{28}$Si:P-$10^{15}$ at 9.7 GHz. Two traces for $^{28}$Si:P-$10^{15}$ were measured as marked: by averaging of the phased echo signal (using conventional quadrature detection) and by averaging the echo magnitude (i.e. disregarding the phase of the echo signal). The faster decay seen in the "quadrature detection" approach results from non-ideal characteristics of the pulse EPR

spectrometer (phase fluctuations of the microwave source and/or fluctuations of the external magnetic field).

Fig. 3. Demonstration of the instantaneous diffusion contribution to $T_2$ in $^{28}$Si:P-10$^{15}$, at 9.7 GHz and temperatures 8.1 and 6.9 K. $1/T_2$ (solid symbols) is plotted as a function of the turning angle ($\theta_2$) of the second microwave pulse in a 2-pulse ESE experiment. Open symbols on the y-axis indicate $1/T_1$ at the respective temperatures. The slope of the linear fit (dashed lines) is proportional to the P concentration and the intercept corresponds to the intrinsic $T_2$ of an isolated donor-electron spin.

Figure 1. Tyryshkin et al.

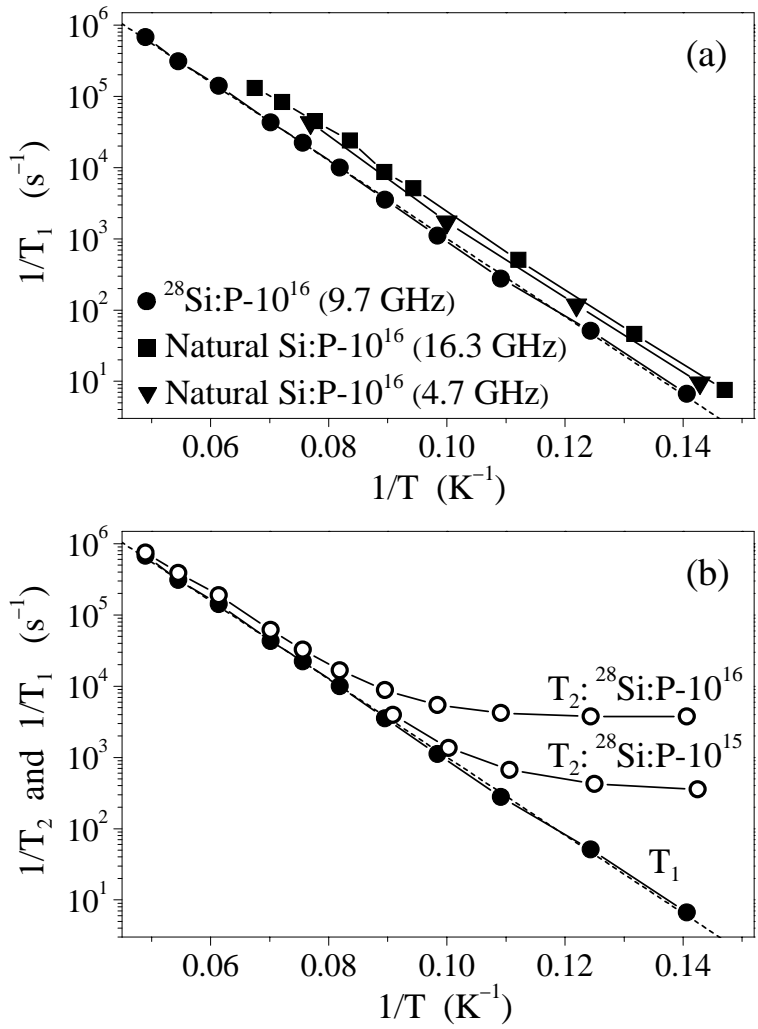

Figure 2. Tyryshkin et al.

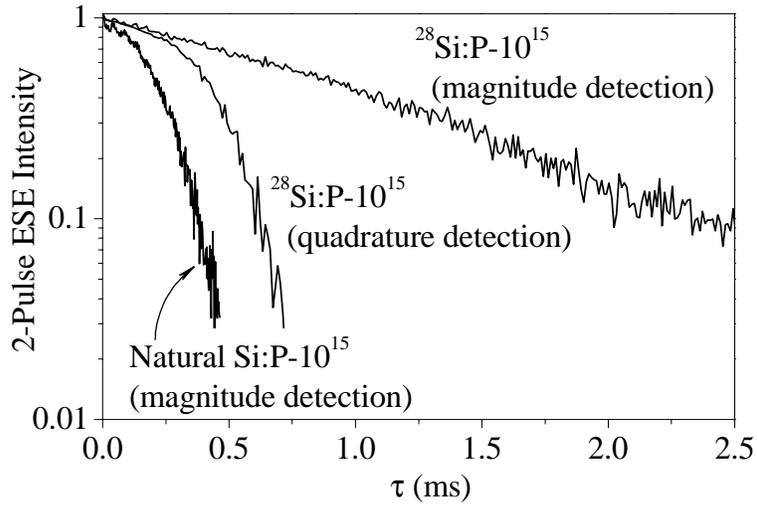

Figure 3. Tyryshkin et al.

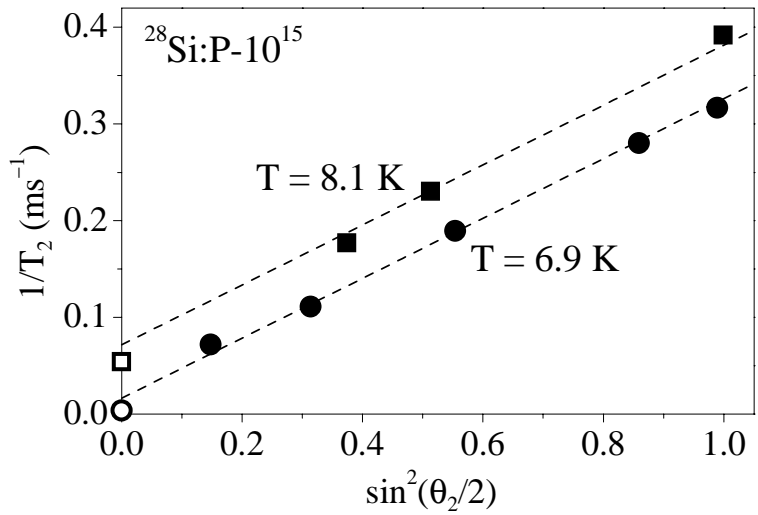